\documentclass[preprint,prd,superscriptaddress,showpacs,nofootinbib,nobibnotes]{revtex4}
\usepackage{amsfonts,bm,natbib}

\usepackage{color}

\newcommand{\be}{\begin{equation}}
\newcommand{\bea}{\begin{eqnarray}}
\newcommand{\ee}{\end{equation}}
\newcommand{\eea}{\end{eqnarray}}

\def\1eq#1{Eq.~(\ref{#1})}
\def\2eqs#1#2{Eqs.~(\ref{#1}) and~(\ref{#2})}
\def\3eqs#1#2#3{Eqs.~(\ref{#1}),~(\ref{#2}) and~(\ref{#3})}
\def\noeq#1{(\ref{#1})} 

\def\G{\Gamma}
\def\genV{V}

\def\diff{\mathrm{d}}

\begin{document}

\title{The Background Field Method as a Canonical Transformation}
\date{March 23, 2012}

\author{D. Binosi}
\affiliation{European Centre for Theoretical Studies in Nuclear
  Physics and Related Areas (ECT*) \\and Fondazione Bruno Kessler,  Villa Tambosi, Strada delle
  Tabarelle 286, 
 I-38123 Villazzano (TN)  Italy}
 
\author{A. Quadri}
\affiliation{Dip. di Fisica, Universit\`a degli Studi di Milano via Celoria 16, I-20133 Milano, Italy\\
and INFN, Sezione di Milano, via Celoria 16, I-20133 Milano, Italy}

\begin{abstract}
\noindent We construct explicitly the canonical transformation that controls the full dependence (local and non-local) of the vertex functional of a Yang-Mills theory on a background field. After  showing  that the canonical transformation found is nothing but a direct field-theoretic generalization of the Lie transform of classical analytical mechanics, we comment on a number of possible applications, and in particular the non perturbative implementation of the background field method on the lattice, the background field formulation of the two particle irreducible formalism, and, finally, the formulation of the Schwinger-Dyson series in the presence of topologically non-trivial configurations.
\end{abstract}


\maketitle


\noindent {\it Introduction.}
Quantization of non-Abelian gauge theories around  
background field configurations is a subject of considerable
interest. 
Since the pioneering work of 't~Hooft~\cite{Hooft:1976fv} 
the path-integral around topologically non-trivial field configurations
has  been used in many different applications, ranging from
chiral soliton models in effective approaches to low-energy QCD \cite{Meier:1996ng} to the study of critical solitons in
supersymmetric models~\cite{Shifman:2007ce}.

Quantum fluctuations around the classical background can be treated
perturbatively by fixing a gauge while retaining
explicit (background) gauge invariance. This is 
the so-called background field method (BFM)~\cite{Abbott:1981ke},
which has been widely used to simplify, technically as well as conceptually, calculations in gauge theories.

While within perturbation theory the BFM has been extended to all orders  both in the continuum~\cite{Abbott:1980hw} and on the lattice~\cite{Luscher:1995vs}, there is yet no clear-cut prescription on how to handle the BFM quantization in non-perturbative approaches  to non-Abelian gauge theories, like, {\it e.g.}, their {\em non-perturbative} formulation on the lattice~\cite{Wilson:1974sk} or their treatment in the 2 particle irreducible (2-PI) formalism of Cornwall, Jackiw and Toumbulis~\cite{Cornwall:1974vz}. One of the open issues in realizing this program is the existence of Gribov copies~\cite{Gribov:1977wm} which prevents a direct generalization of the perturbative treatment. 

Clearly, if one were able to implement the BFM for {\em both}
non-perturbative lattice gauge theory and the 2-PI formalism, one could
make significant progress in the computation of non-perturbative lattice
quantities, as well as in understanding the matching with their
continuum counterparts.
For instance, the one-loop
correspondence \cite{Denner:1994nn} between the pinch technique \cite{Cornwall:1981zr} and the BFM
 Green's functions has been shown to
hold true to all orders for the background dependent
amplitudes \cite{Papavassiliou:1994yi,Binosi:2002ft}.
Then the simulation of the background gluon two-point
function 
on a lattice gauge fixed in the BFM (Feynman) gauge would allow one to construct a renormalization group invariant propagator (that is independent of the renormalization point $\mu$ chosen, or, conversely, of the lattice space chosen), by simply
multiplying the unrenormalized propagator by the square of the unrenormalized charge, as in QED. 
Moreover, the ability to extend the BFM to the aforementioned contexts would open up a wide range of gauge-invariant simulations and variational estimates. This would translate into very useful phenomenology for addressing the properties of the infrared sector of Yang-Mills theories and in particular phenomena like confinement, chiral symmetry breaking and/or dynamical gluon mass generation.

Surprisingly enough, it turns out that one can give
a very simple characterization of the dependence of the
effective action on the background that holds in a very
general setting. Specifically, 
as we will show in this letter, it turns out that, whenever the extended Slavnov Taylor (ST) identity in the presence of the background is fulfilled, the
background dependence of the effective action is governed
by a canonical transformation with respect to (w.r.t.) the fundamental
Batalin-Vilkovisky (BV) bracket of the underlying gauge theory. 
Consequently, one can draw a fruitful analogy with the theory of Lie 
transforms in classical analytical mechanics, and obtain simple and powerful formulas for the finite canonical
transformation that fix (uniquely) the background-dependent amplitudes in terms of those at zero background.

In a purely nonperturbative setting, the definition of the ST identity is a delicate issue requiring a careful analysis, that exceeds  the purpose and scope of this letter.
Nevertheless, we would like to point out that the approach proposed in the present paper does not require the presence of dynamical ghosts. This represents an advantage w.r.t. the conventional techniques for implementing the BFM on the lattice, since, as we will explicitly see, it allows in principle to evade the Neuberger's 0/0 problem~\cite{Neuberger:1986xz}. 

\bigskip

\noindent {\it Background fields and canonical transformations.}
Within the BV framework, the {\it complete} vertex functional $\Gamma$ of a SU(N) Yang-Mills theory, quantized in a linear background gauge, can be written in terms of the fields $\phi=(A^a_\mu,c^a,\bar{c}^a,b^a)$, the antifields $\phi^*=(A^{*a}_\mu,c^{*a})$, the background field $\widehat{A}^a_\mu$ and its associated source $\Omega^a_\mu$~\cite{Grassi:1995wr}. The antifields $(\bar c^{*a}, b^{*a})$ are not needed, since the fields  $b^a$ and $\bar c^a$ form a BRST doublet~\cite{Barnich:2000zw,Quadri:2002nh}, {\it i.e.}, a set of variables $u,v$
transforming under the BRST differential $s$ according to
 $su = v$, $sv=0$. This, together with the fact that the $b$-dependence is confined at tree-level, allows one to eliminate the doublet $(\bar{c}^a,b^a)$ by means of a canonical transformation yielding the so-called {\it reduced} functional. 

For example, if one considers the background Lorentz-covariant gauge-fixing function $\widehat{\cal F}^a=[\widehat {\cal D}^\mu (A-\widehat A)_\mu]^a$ with $\widehat{\cal D}^{ab}_\mu=\delta^{ab}\partial _\mu +f^{acb}\widehat{A}^c_\mu$ the background covariant derivative, the complete tree-level vertex functional reads
\bea
\G^{(0)}&=&\int\!\mathrm{d}^4x\left[-\frac14F^a_{\mu\nu} F^{a\mu\nu}-\bar{c}^a(\widehat{\cal D}_\mu{\cal D}^\mu c)^a-({\cal D}^\mu \bar{c})^a\Omega^a_\mu-\frac\xi2(b^a)^2+b^a[\widehat{\cal D}^\mu(A-\widehat{A})_\mu]^a\right.\nonumber\\
&+&\left.A^{*a}_\mu\left({\cal D}^\mu c\right)^a+\frac12f^{abc} c^{*a}c^bc^c\right].
\label{tlvf}
\eea
The reduced functional is then obtained by first defining
$$
\widetilde \G = \G - \int\!\diff^4x \, b^a [\widehat {\cal D}^\mu (A-\widehat A)_\mu]^a+\frac\xi2\int\!\diff^4x\,(b^a)^2,
$$
and then eliminating $\bar c^a$  through the antifield redefinition $\widetilde A^{*a}_\mu = A^{*a}_{\mu} + (\widehat {\cal D}_\mu \bar c)^a$, which, due to the antighost equation $\frac{\delta\Gamma}{\delta {\bar{c}^a}}=-\widehat{\cal D}^{ab}_\mu\frac{\delta\Gamma}{\delta {A^{*b}_\mu}}+({\cal D}^\mu \Omega_\mu)^a$, represents the only combination through which  the vertex functional could possibly depend on $\bar c^a$. In what follows we will always use the reduced vertex functional and thus drop the tilde symbols on all quantities.  

As shown in~\cite{Binosi:2012pd} the {\it extended} ST identity in the presence of a background field can be written as
\be
\int\!\diff^4x\, \Omega^a_\mu(x)
\frac{\delta \G}{\delta \widehat A^a_\mu(x)} = 
- \frac{1}{2}\, \{\G,\G\} .
\label{m.1}
\ee
where $\{X,Y\}$ represents the BV bracket defined as (only left derivative assumed in what follows)~\cite{Gomis:1994he}
\bea
\{X,Y\} &=& \int\!\diff^4x \sum_\phi
\left[ (-1)^{\epsilon_{\phi} (\epsilon_X+1)}
\frac{\delta X}{\delta \phi} \frac{\delta Y}{\delta \phi^*}
-(-1)^{\epsilon_{\phi^*} (\epsilon_X+1)}
\frac{\delta X}{\delta \phi^*} \frac{\delta Y}{\delta \phi}
\right].
\label{BVbracket}
\eea
The sum runs over the fields $\phi = (A^a_\mu,c^a)$ and the 
corresponding antifields 
$\phi^* = (A^{*a}_\mu, c^{*a})$, with $\epsilon_\phi$, $\epsilon_{\phi^*}$ and $\epsilon_X$ representing the statistics of the field $\phi$, the antifield $\phi^*$ and the functional $X$ respectively. For the graded properties of the BV bracket the reader is referred to~\cite{Gomis:1994he}.

If one now takes the derivative of \1eq{m.1} w.r.t. $\Omega^a_\mu$ and set the latter source equal to zero afterwards, the resulting equation~\cite{Binosi:2012pd}
\bea
\left.\frac{\delta \G}{\delta \widehat A^a_\mu(x)}\right|_{\Omega=0} = 
\left.- \{ \frac{\delta \G}{\delta \Omega^a_\mu(x)},
\G \}\right|_{\Omega=0},
\label{m.2}
\eea
shows that the 
derivative of the vertex functional w.r.t. 
the background field equals the 
effect of an infinitesimal canonical
transformation (w.r.t. the BV bracket)
on the vertex functional itself. Then,
since the BV bracket does not depend
on either $\widehat A^a_\mu$ or $\Omega^a_\mu$, if one were able to write the finite
canonical transformation generated by the fermion 
$\Psi^a_\mu(x)=\frac{\delta \G}{\delta \Omega^a_\mu(x)}$, one would control the full dependence of $\G$
on the background fields; and this would happen not only at the level of the counterterms  of $\G$, but rather for the full 1-PI Green's functions, thus giving control even over the non-local dependence on the background.  

The problem can be thus stated as follows: given the field and antifield variables $\phi$, $\phi^*$, which are canonical w.r.t. the BV bracket~\noeq{BVbracket}, {\it i.e.},
\bea
\{\phi_i(x),\phi_j(y)\}&=&\{\phi^*_i(x),\phi^*_j(y)\}=0\nonumber \\
\{\phi_i(x),\phi^*_j(y)\}&=&\delta_{ij}\delta^4(y-x),\nonumber
\eea
and the background field $\widehat{A}^a_\mu$,
find the canonical mapping 
$$
(\phi(x),\phi^*(x);\widehat{A}^a_\mu(x))\mapsto(\Phi(x),\Phi^*(x)),
$$
to the new field and antifield variables $\Phi$ and $\Phi^*$
such that the ST identity~\noeq{m.2} written in these new variables is automatically satisfied. 
This last condition translates in a relatively straightforward fashion, 
into determining the canonical variables $\Phi$ and $\Phi^*$ which are also solutions of the two equations
\bea
\frac{\delta\Phi(y)}{\delta \widehat{A}^a_\mu(x)}&=&\frac{\delta\Psi^a_\mu(x)}{\delta \Phi^*(y)}=\{\Phi(y),\Psi^a_\mu(x)\},\nonumber \\
\frac{\delta\Phi^*(y)}{\delta \widehat{A}^a_\mu(x)}&=&-\frac{\delta\Psi^a_\mu(x)}{\delta \Phi(y)}=\{\Phi^*(y),\Psi^a_\mu(x)\}.
\label{STcond}
\eea

Before proceeding to construct explicitly the canonical mapping, let us notice that a (recursive) solution of the {\it finite} canonical transformation has been already derived by means of homotopy techniques in~\cite{Binosi:2012pd}, where it was found that this solution fails to respect the (naively expected) exponentiation pattern, due to the dependence of the generating functional $\Psi^a_\mu$ on the background field~$\widehat{A}^a_\mu$.

In order to find the explicit canonical transformation, let us then introduce the operator 
$$
\Delta_{\Psi^{a}_\mu(x)}=\{\cdot,\Psi^{a}_\mu(x)\}+\frac\delta{\delta \widehat{A}^a_\mu(x)}.
$$
The first term above represents a (graded) generalization (to the BV bracket and a fermionic generator) of the classical Lie derivative w.r.t a (bosonic) generator (in which case the bracket would be the usual Poisson bracket); the second term takes into account the above observation on the exponentiation failure.

Using then the properties of the BV bracket, it is not particularly difficult to establish the following relations 
\bea
& & \Delta_{\Psi^a_\mu(x)} (\alpha X + \beta Y) =
\alpha \Delta_{\Psi^a_\mu(x)} X + \beta
\Delta_{\Psi^a_\mu(x)} Y, \nonumber \\
& &  \Delta_{\Psi^a_\mu(x)} (XY) = 
X \Delta_{\Psi^a_\mu(x)} Y + (-1)^{\epsilon_X \epsilon_Y} Y \Delta_{\Psi^a_\mu(x)} X,  \nonumber\\
& & \Delta_{\Psi^a_\mu(x)} \{ X, Y \} = 
 \{  \Delta_{\Psi^a_\mu(x)} X, Y \}
+ \{  X,  \Delta_{\Psi^a_\mu(x)} Y \}.\nonumber
\eea
The first two equations above establish that  $\Delta_{\Psi^a_\mu(x)}$ gives rise to a graded derivation with the usual statistics, while the last formula allows us to determine the important result 
\bea
\int_1\!\cdots\int_n\,
\widehat A_1 \cdots\hat A_n\,
\Delta_{\Psi_n} \cdots \Delta_{\Psi_1}
\{ X, Y \} &=&
\sum_{0 \leq m \leq n}
\pmatrix{n \cr m} \{
\Delta_{\Psi_1} \cdots \Delta_{\Psi_m} X, 
\Delta_{\Psi_{m+1}}
\dots \Delta_{\Psi_n}Y \},\nonumber\\
\label{20n}
\eea
where we have introduced the shorthand notation \mbox{$\int_i=\int\diff^4y_i$}, \mbox{$\widehat{A}_i=\widehat{A}^{a_i}_{\mu_i}(y_i)$} and $\Psi_i=\Psi^{a_i}_{\mu_i}(y_i)$. 

From the operator $\Delta_\Psi$ one can then define a mapping $E_{\Psi}$ given in terms of a formal power series in the background field $\widehat{A}$ as follows
\bea
\Phi(x) &=& E_{\Psi}(\phi(x)) 
\label{cantr}\\
&\equiv& \sum_{n \geq 0}
\frac{1}{n!}  \int_1\! \cdots\! \int_n\!
\widehat A_1\cdots\widehat A_n  
\left [ \Delta_{\Psi_n}\! \cdots \Delta_{\Psi_1} \phi(x) \right ]_{\widehat A=0}, \nonumber 
\eea
with an identical expansion holding for the antifields variables. Then~\1eq{cantr} constitutes the sought for canonical mapping between the old and the new variables.  

Indeed, on the one hand, the canonicity property is a direct consequence of \1eq{20n} above, since the latter directly implies the identity $E_{\Psi}\{X,Y\}=\{E_{\Psi}X,E_{\Psi}Y\}$.
On the other hand, to see that the new variables are indeed solutions of Eqs.~(\ref{STcond}), let us concentrate on the case of a bosonic field $\Phi$ and expand both the latter and the fermionic generator $\Psi^a_\mu$ in power series w.r.t. the background field $\widehat{A}$. Schematically, one has
\bea
\Phi&=&\phi+\sum_{n \geq 0}
\frac{1}{n!}  \int_1\! \cdots\! \int_n\!
\widehat A_1\cdots\widehat A_n \Phi_{1\cdots n},\nonumber \\
\Psi_0&=&\psi_0+\sum_{n \geq 0}
\frac{1}{n!}  \int_1\! \cdots\! \int_n\!
\widehat A_1\cdots\widehat A_n \Psi_{01\cdots n},\nonumber
\eea
and finds up to third order in $\widehat{A}$ 
\bea
& & \Delta_{\Psi_1}\phi\vert_{\widehat{A}=0}=\{\phi,\psi_1\},\nonumber \\
& & \Delta_{\Psi_2}\Delta_{\Psi_1}\phi\vert_{\widehat{A}=0}=\{\{\phi,\psi_1\},\psi_2\}+\{\phi,\Psi_{12}\},\nonumber 
\\
& & \Delta_{\Psi_3}\Delta_{\Psi_2}\Delta_{\Psi_1}\phi\vert_{\widehat{A}=0}=\{\{\{\phi,\psi_1\},\psi_2\},\psi_3\}+\{\phi,\Psi_{123}\}\nonumber 
\\
&& + \{\{\phi,\Psi_{12}\},\psi_3\}+ \{\{\phi,\Psi_{23}\},\psi_1\}+ \{\{\phi,\Psi_{31}\},\psi_2\},\nonumber 
\eea
where in the last equation we have symmetrized all indices, and used the (graded) Jacobi identity together with the result $
\int_1\!\int_2\!\widehat{A}_1\widehat{A}_2\frac{\delta}{\delta\widehat{A}_3}\{\Psi_1,\Psi_2\}=0.
$
It can then be  checked that the above terms are indeed the solutions (up to third order in $\widehat{A}$) of the first of Eqs.~(\ref{STcond}). The fermionic case, {\it e.g.},  a fermionic antifield $\Phi^*$, can be treated in exactly the same way.

\bigskip

\noindent {\it Discussion.} There are 
several comments that can be made w.r.t. the canonical transformation~(\ref{cantr}). 

To begin with, it should be noticed that such transformation, together with the method used for constructing it, is nothing but a direct generalization of the procedure developed long ago
by Deprit~\cite{Deprit:1969aa}, to construct canonical mappings in the form of (formal) power series in a small parameter~$\epsilon$, in cases where the generating function itself  explicitly depends on $\epsilon$. In this case, the problem one tries to address is the following: given a function $\genV$ that depends on the (canonical) variables $q$, $p$ and a parameter $\epsilon$, find a canonical mapping
$(q,p;\epsilon)\mapsto(Q,P)$,
such that the new variables $Q$ and $P$ satisfy the equations
$$
\frac{\diff Q}{\diff\epsilon}=\frac{\partial}{\partial P}\genV (p,q;\epsilon);\qquad
\frac{\diff P}{\diff\epsilon}=-\frac{\partial}{\partial Q}\genV (p,q;\epsilon).
$$

The solution to this problem is found~\cite{Deprit:1969aa} by enlarging the concept of a Lie series through the introduction of  the operator
$\Delta_\genV=\{\cdot, V \}+\frac{\partial}{\partial\epsilon}$
(the bracket being now the usual Poisson bracket) and next  defining, for any given function $f$ of the variables $p$, $q$ and $\epsilon$,  the formal power series (Lie transform generated by $\genV$)
$$
E_\genV(f)=\sum_n\frac1{n!}\epsilon^n\Delta^n_{\genV(q,p;\epsilon)}f(q,p;\epsilon)\vert_{\epsilon=0}.
$$
Then, the new canonical variables are given by the Lie transforms
$Q=E_\genV(q)$ and $P=E_\genV(p)$.
We thus  see that the canonical transformation found has a sound geometric interpretation, being a direct field-theoretic generalization of the Lie transform in classical analytical mechanics.
Given this connection, one might conversely wonder what advantages could bring the use of the homotopy techniques of~\cite{Binosi:2012pd} in a completely classical context.

Second, the mapping~(\ref{cantr}) provides a new set of field variables such that when the conventional Green's functions are written in terms of these new variables they would coincide with those calculated in the BFM, thus explaining the aforementioned correspondence between them. It would be then very interesting to supplement the current formulation of the canonical transformation with Nielsen identities~\cite{Nielsen:1975fs}  and study the flow of~(\ref{cantr}) as $\xi$ moves towards the critical value $\xi=1$, where it is known that the BFM Green's functions acquire additional physical properties~\cite{Binosi:2009qm}.

Third, it should be noticed that at no point in this analysis we have relied on the Ward identity usually associated with  background linear gauge fixings such as the one used for illustrative purposes in~(\ref{tlvf}). Indeed, the only requirement we have on the gauge fixing fermion is that it is possible to construct the canonical mapping that eliminates the BRST doublet $(\bar{c}^a,b^a)$, thus allowing for the writing of the reduced vertex functional and ultimately of  the extended ST identity~\noeq{m.1} (which happens in the vast majority of cases).
This shows that it is the (extended) ST identity and not the Ward identity that forces the strongest constraints on the theory, in agreement with the findings of~\cite{Ferrari:2000yp}, where it was shown that the background Ward identity alone is not able to guarantee physical unitarity ({\it i.e.}, the cancellation of the intermediate ghost states in physical amplitudes) in the absence of the ST identity.

As for applications we can think of at least three.
The first one is in relation to the non-perturbative formulation of the BFM on the lattice. At first sight this claim looks surprising, since such a formulation would require a BRST invariant integration over link variables, and it is well known that the so-called Neuberger $0/0$ problem~\cite{Neuberger:1986xz}  forbids a direct non perturbative generalization of the BRST symmetry. 
Thus the extended ST identity, 
which clearly constitutes  the central pillar of our construction, would not be present either. However, 
notice that the canonical transformation~\noeq{cantr} 
can also be written in a (gauge-invariant) model where
the ghosts are replaced by external classical
anticommuting sources, {\it i.e.},  the BV bracket spans the gauge field $A^a_\mu$ and its antifield $A^{*a}_\mu$ and, for a gauge fixing obtained by minimizing some functional $F[g]$ over the gauge group, the parameters of the group element $g$ and their antifields. 
Thus the fact that {\em dynamical} ghosts need not be present in the formulation, overcomes the  absence at the non-perturbative level of the BRST symmetry. Assume then that one is able to fix a background gauge, {\it e.g.}, through the minimization of a suitable functional $F[g]$ (recall that our derivation does not rely on the particular gauge fixing chosen); assume also that at a {\it fixed} background $\widehat{A}$ such functional depends on the gauge field only through the combination $A^g_\mu - \widehat A_\mu$ (leading to the most economical generalization of the ordinary Landau gauge functional) with $A_\mu^g = g^\dagger  A_\mu g - i \partial_\mu g^\dagger g$ and $g$ a gauge group element. 
The simplest generalization of the ordinary Landau gauge functional
is
\bea
F[g] = - \int \mathrm{d}^4x \, \mathrm{Tr} ( A^g_\mu - \widehat A_\mu)^2 \, .
\label{bkg.functional}
\eea
When minimized, it gives the background Landau gauge
condition $\widehat {\cal D}_\mu(A^g_\mu-\widehat A_\mu)=0$\footnote{ Notice that the background Landau gauge condition for the
quantum field $Q_\mu$ can be obtained by finding the extrema 
of the functional $\int \mathrm{d}^4x \, \mathrm{Tr}  ~ Q_\mu \widehat A^g_\mu$, where 
$\widehat A^g_\mu$ is the gauge-transformed background field, 
{\it i.e.}, one  gauge-rotates the background field by keeping $Q_\mu$ fixed.
We remark however that this procedure does not select
in general a unique representative along the gauge orbit of
$Q_\mu$.}.
Then {\it on the minimum} of the functional~(\ref{bkg.functional}) the mapping 
\be
A\to A^g(A,\widehat{A})-\widehat{A},
\label{discrmap}
\ee 
defines the action of the canonical transformation on the gauge field, thus generalizing non perturbatively the background quantum splitting. 
In the presence of Gribov copies multiple minima exist that are parametrized by different functions $g_i(A, \widehat{A})$; however, the canonical mapping~(\ref{discrmap}) allows for reconstructing the full dependence on the background also in such case, provided that we restrict ourselves to the region of validity of each $g_i$.

At this point, the strategy would then be to reconstruct the dependence on the background of the various quantities calculated through this canonical transformation (with a suitable extension to the gauge antifield). Notice that this canonical mapping provides highly non-trivial constraints, relating quantum and background Green's
functions, that can therefore be tested, at least in principle, on the lattice.

A second application is the BFM formulation of the 2-PI formalism~\cite{Cornwall:1974vz}. To get an idea of how this can be accomplished, observe that the extended ST identity~\noeq{m.1} can be rewritten in terms of the generator of the connected Green's functions $W[J]=\Gamma[\Phi]+\int\!J\Phi$ as
\be
\int\!\diff^4x\, \Omega^a_\mu(x)
\frac{\delta W}{\delta \widehat A^a_\mu(x)} = 
-\int\!\diff^4x\,J(x)\frac{\delta W}{\delta\Phi^*(x)}.
\label{1PI}
\ee
Starting from the connected diagrams, which are assumed to satisfy Eq.~(\ref{1PI}), one can perform a double Legendre transform 
$$
W[J,K]=\Gamma[\Phi,G]+\int\!J\Phi+1/2\int\!\!\int\!\Phi K\Phi+1/2\hbar\int\!\!\int\!GK,
$$ 
and derive the corresponding extended ST identity for the 2-PI effective action 
$\Gamma[\Phi,G]$. 
This however implies the introduction of the BRST doublet
$s\chi=\pm K$ and $sK=0$ 
(the $\pm$ sign corresponding to bosonic/fermionic fields) and the addition to the tree-level action of the composite operator term
$$
s\frac12\int\!\!\!\int\!\Phi \chi\Phi=\frac12\int\!\!\!\int\!\Phi K\Phi+\int\!\!\!\int\!\chi\frac{\delta\Gamma}{\delta\Phi^*}\Phi.
$$
Due to the nilpotency of the BRST operator this term does not violate the ST identity; then, one can study the extra terms that are bound to appear in~\1eq{1PI} and at the same time keep under control the renormalization of the operators added.  

The third application is the study of the conditions under which the (non-perturbative) Schwinger-Dyson (SD) equations can be reliably trusted when expanding around non-trivial vacua. As shown in~\cite{Szczepaniak:2011bs} for some toy models, a naive SD expansion is poor when the potential admits more than one minimum; on the other hand, a modified SD formulation, is required in order to improve on the saddle point approximation in such cases. The formalism developed here and in~\cite{Binosi:2012pd} can indeed help in formulating the SD expansion in the presence of topologically non trivial vacuum configurations (instantons, center vortices, monopoles, etc.); indeed the framework of~\cite{Binosi:2007pi} and its related truncation scheme, which in the Landau gauge compares favorably with large-volume lattice simulations, could be generalized to study the effects due to the presence of  such solitons in the theory vacuum.
 
Concluding, in this letter we have explicitly constructed the canonical transformation that controls the full dependence of the vertex functional $\Gamma$ on the background field~$\widehat{A}^a_\mu$. 
Though being an interesting result in its own right, especially given its connection with classical analytical mechanics, its strongest appeal resides in the many interesting directions it opens up.

\bigskip

\noindent{\it Acknowledgments} Useful conversations with J. Cornwall, A. Cucchieri, T. Mendes, J. Papavassiliou, A. Szczepaniak and A. Slavnov  are gratefully acknowledged.


\begin{thebibliography}{23}
\expandafter\ifx\csname natexlab\endcsname\relax\def\natexlab#1{#1}\fi
\expandafter\ifx\csname bibnamefont\endcsname\relax
  \def\bibnamefont#1{#1}\fi
\expandafter\ifx\csname bibfnamefont\endcsname\relax
  \def\bibfnamefont#1{#1}\fi
\expandafter\ifx\csname citenamefont\endcsname\relax
  \def\citenamefont#1{#1}\fi
\expandafter\ifx\csname url\endcsname\relax
  \def\url#1{\texttt{#1}}\fi
\expandafter\ifx\csname urlprefix\endcsname\relax\def\urlprefix{URL }\fi
\providecommand{\bibinfo}[2]{#2}
\providecommand{\eprint}[2][]{\url{#2}}

\bibitem[{\citenamefont{'t~Hooft}(1976)}]{Hooft:1976fv}
\bibinfo{author}{\bibfnamefont{G.}~\bibnamefont{'t~Hooft}},
  \bibinfo{journal}{Phys.Rev.} \textbf{\bibinfo{volume}{D14}},
  \bibinfo{pages}{3432} (\bibinfo{year}{1976}).

\bibitem[{\citenamefont{Meier and Walliser}(1997)}]{Meier:1996ng}
\bibinfo{author}{\bibfnamefont{F.}~\bibnamefont{Meier}} \bibnamefont{and}
  \bibinfo{author}{\bibfnamefont{H.}~\bibnamefont{Walliser}},
  \bibinfo{journal}{Phys.Rept.} \textbf{\bibinfo{volume}{289}},
  \bibinfo{pages}{383} (\bibinfo{year}{1997}).

\bibitem[{\citenamefont{Shifman and Yung}(2007)}]{Shifman:2007ce}
\bibinfo{author}{\bibfnamefont{M.}~\bibnamefont{Shifman}} \bibnamefont{and}
  \bibinfo{author}{\bibfnamefont{A.}~\bibnamefont{Yung}},
  \bibinfo{journal}{Rev.Mod.Phys.} \textbf{\bibinfo{volume}{79}},
  \bibinfo{pages}{1139} (\bibinfo{year}{2007}).

\bibitem[{\citenamefont{Abbott}(1982)}]{Abbott:1981ke}
\bibinfo{author}{\bibfnamefont{L.~F.} \bibnamefont{Abbott}},
  \bibinfo{journal}{Acta Phys. Polon.} \textbf{\bibinfo{volume}{B13}},
  \bibinfo{pages}{33} (\bibinfo{year}{1982}), and references therein.

\bibitem[{\citenamefont{Abbott}(1981)}]{Abbott:1980hw}
\bibinfo{author}{\bibfnamefont{L.~F.} \bibnamefont{Abbott}},
  \bibinfo{journal}{Nucl. Phys.} \textbf{\bibinfo{volume}{B185}},
  \bibinfo{pages}{189} (\bibinfo{year}{1981}).

\bibitem[{\citenamefont{Luscher and Weisz}(1995)}]{Luscher:1995vs}
\bibinfo{author}{\bibfnamefont{M.}~\bibnamefont{Luscher}} \bibnamefont{and}
  \bibinfo{author}{\bibfnamefont{P.}~\bibnamefont{Weisz}},
  \bibinfo{journal}{Nucl.Phys.} \textbf{\bibinfo{volume}{B452}},
  \bibinfo{pages}{213} (\bibinfo{year}{1995}).

\bibitem[{\citenamefont{Wilson}(1974)}]{Wilson:1974sk}
\bibinfo{author}{\bibfnamefont{K.~G.} \bibnamefont{Wilson}},
  \bibinfo{journal}{Phys. Rev.} \textbf{\bibinfo{volume}{D10}},
  \bibinfo{pages}{2445} (\bibinfo{year}{1974}).

\bibitem[{\citenamefont{Cornwall et~al.}(1974)\citenamefont{Cornwall, Jackiw,
  and Tomboulis}}]{Cornwall:1974vz}
\bibinfo{author}{\bibfnamefont{J.~M.} \bibnamefont{Cornwall}},
  \bibinfo{author}{\bibfnamefont{R.}~\bibnamefont{Jackiw}}, \bibnamefont{and}
  \bibinfo{author}{\bibfnamefont{E.}~\bibnamefont{Tomboulis}},
  \bibinfo{journal}{Phys. Rev.} \textbf{\bibinfo{volume}{D10}},
  \bibinfo{pages}{2428} (\bibinfo{year}{1974}).


\bibitem{Gribov:1977wm} 
  V.~N.~Gribov,
  Nucl.\ Phys.\  {\bf B139}, 1 (1978).

\bibitem{Denner:1994nn}
  A.~Denner, G.~Weiglein and S.~Dittmaier,
  Phys.\ Lett.\  {\bf B333},  420 (1994).


\bibitem[{\citenamefont{Cornwall}(1982)}]{Cornwall:1981zr}
\bibinfo{author}{\bibfnamefont{J.~M.} \bibnamefont{Cornwall}},
  \bibinfo{journal}{Phys. Rev.} \textbf{\bibinfo{volume}{D26}},
  \bibinfo{pages}{1453} (\bibinfo{year}{1982}).

\bibitem{Papavassiliou:1994yi} 
  J.~Papavassiliou,
  Phys.\ Rev.\  {\bf D51}, 856 (1995).


\bibitem[{\citenamefont{Binosi and Papavassiliou}(2002)}]{Binosi:2002ft}
\bibinfo{author}{\bibfnamefont{D.}~\bibnamefont{Binosi}} \bibnamefont{and}
  \bibinfo{author}{\bibfnamefont{J.}~\bibnamefont{Papavassiliou}},
  \bibinfo{journal}{Phys. Rev.} \textbf{\bibinfo{volume}{D66}},
  \bibinfo{pages}{111901(R)} (\bibinfo{year}{2002}); 
  \bibinfo{journal}{J. Phys.} \textbf{\bibinfo{volume}{G30}},
  \bibinfo{pages}{203} (\bibinfo{year}{2004}).

\bibitem[{\citenamefont{Neuberger}(1987)}]{Neuberger:1986xz}
\bibinfo{author}{\bibfnamefont{H.}~\bibnamefont{Neuberger}},
  \bibinfo{journal}{Phys.Lett.} \textbf{\bibinfo{volume}{B183}},
  \bibinfo{pages}{337} (\bibinfo{year}{1987}).

\bibitem[{\citenamefont{Grassi}(1996)}]{Grassi:1995wr}
\bibinfo{author}{\bibfnamefont{P.~A.} \bibnamefont{Grassi}},
  \bibinfo{journal}{Nucl. Phys.} \textbf{\bibinfo{volume}{B462}},
  \bibinfo{pages}{524} (\bibinfo{year}{1996}).

\bibitem[{\citenamefont{Barnich et~al.}(2000)\citenamefont{Barnich, Brandt, and
  Henneaux}}]{Barnich:2000zw}
\bibinfo{author}{\bibfnamefont{G.}~\bibnamefont{Barnich}},
  \bibinfo{author}{\bibfnamefont{F.}~\bibnamefont{Brandt}}, \bibnamefont{and}
  \bibinfo{author}{\bibfnamefont{M.}~\bibnamefont{Henneaux}},
  \bibinfo{journal}{Phys.Rept.} \textbf{\bibinfo{volume}{338}},
  \bibinfo{pages}{439} (\bibinfo{year}{2000}).

\bibitem[{\citenamefont{Quadri}(2002)}]{Quadri:2002nh}
\bibinfo{author}{\bibfnamefont{A.}~\bibnamefont{Quadri}},
  \bibinfo{journal}{JHEP} \textbf{\bibinfo{volume}{0205}}, \bibinfo{pages}{051}
  (\bibinfo{year}{2002}).

\bibitem[{\citenamefont{Binosi and Quadri}(2012)}]{Binosi:2012pd}
\bibinfo{author}{\bibfnamefont{D.}~\bibnamefont{Binosi}} \bibnamefont{and}
  \bibinfo{author}{\bibfnamefont{A.}~\bibnamefont{Quadri}}
  (\bibinfo{year}{2012}), \eprint{1201.1807}.

\bibitem[{\citenamefont{Gomis et~al.}(1995)\citenamefont{Gomis, Paris, and
  Samuel}}]{Gomis:1994he}
\bibinfo{author}{\bibfnamefont{J.}~\bibnamefont{Gomis}},
  \bibinfo{author}{\bibfnamefont{J.}~\bibnamefont{Paris}}, \bibnamefont{and}
  \bibinfo{author}{\bibfnamefont{S.}~\bibnamefont{Samuel}},
  \bibinfo{journal}{Phys.Rept.} \textbf{\bibinfo{volume}{259}},
  \bibinfo{pages}{1} (\bibinfo{year}{1995}).

\bibitem[{\citenamefont{Deprit}(1969)}]{Deprit:1969aa}
\bibinfo{author}{\bibfnamefont{A.}~\bibnamefont{Deprit}},
  \bibinfo{journal}{Celest. Mech. Dyn. Astr.} \textbf{\bibinfo{volume}{1}}, \bibinfo{pages}{12}  (\bibinfo{year}{1969}).

\bibitem[{\citenamefont{Nielsen}(1975)}]{Nielsen:1975fs}
\bibinfo{author}{\bibfnamefont{N.~K.} \bibnamefont{Nielsen}},
  \bibinfo{journal}{Nucl. Phys.} \textbf{\bibinfo{volume}{B101}},
  \bibinfo{pages}{173} (\bibinfo{year}{1975}).

\bibitem[{\citenamefont{Binosi and Papavassiliou}(2009)}]{Binosi:2009qm}
\bibinfo{author}{\bibfnamefont{D.}~\bibnamefont{Binosi}} \bibnamefont{and}
  \bibinfo{author}{\bibfnamefont{J.}~\bibnamefont{Papavassiliou}},
  \bibinfo{journal}{Phys.Rept.} \textbf{\bibinfo{volume}{479}},
  \bibinfo{pages}{1} (\bibinfo{year}{2009}).

\bibitem[{\citenamefont{Ferrari et~al.}(2001)\citenamefont{Ferrari, Picariello,
  and Quadri}}]{Ferrari:2000yp}
\bibinfo{author}{\bibfnamefont{R.}~\bibnamefont{Ferrari}},
  \bibinfo{author}{\bibfnamefont{M.}~\bibnamefont{Picariello}},
  \bibnamefont{and} \bibinfo{author}{\bibfnamefont{A.}~\bibnamefont{Quadri}},
  \bibinfo{journal}{Annals Phys.} \textbf{\bibinfo{volume}{294}},
  \bibinfo{pages}{165} (\bibinfo{year}{2001}).

\bibitem[{\citenamefont{Szczepaniak and Reinhardt}(2011)}]{Szczepaniak:2011bs}
\bibinfo{author}{\bibfnamefont{A.~P.} \bibnamefont{Szczepaniak}}
  \bibnamefont{and}
  \bibinfo{author}{\bibfnamefont{H.}~\bibnamefont{Reinhardt}},
  \bibinfo{journal}{Phys.Rev.} \textbf{\bibinfo{volume}{D84}},
  \bibinfo{pages}{056011} (\bibinfo{year}{2011}).

\bibitem[{\citenamefont{Binosi and Papavassiliou}(2008)}]{Binosi:2007pi}
\bibinfo{author}{\bibfnamefont{D.}~\bibnamefont{Binosi}} \bibnamefont{and}
  \bibinfo{author}{\bibfnamefont{J.}~\bibnamefont{Papavassiliou}},
  \bibinfo{journal}{Phys.Rev.} \textbf{\bibinfo{volume}{D77}},
  \bibinfo{pages}{061702} (\bibinfo{year}{2008});
  \bibinfo{journal}{JHEP} \textbf{\bibinfo{volume}{0811}}, \bibinfo{pages}{063}
  (\bibinfo{year}{2008}{\natexlab{b}}).

\end{thebibliography}
\end{document}